\documentstyle [12pt,epsfig]{article} 
\makeatletter

\@addtoreset{equation}{section}
\makeatother

\newcommand{\ba}{\begin{eqnarray}}
\newcommand{\ea}{\end{eqnarray}}

\newcommand{\OP}{{\rm O'}}
\newcommand{\LP}{{\rm L'}}
\newcommand{\LA}{{\rm L}}
\newcommand{\PP}{{\rm P'}}
 
\begin{document}
\newcommand{\BS}{\bigskip}
\newcommand{\SECTION}[1]{\BS{\large\section{\bf #1}}}
\newcommand{\SUBSECTION}[1]{\BS{\large\subsection{\bf #1}}}
\newcommand{\SUBSUBSECTION}[1]{\BS{\large\subsubsection{\bf #1}}}

\begin{titlepage}
\begin{center}
\vspace*{2cm}
{\large \bf A sign error in the Minkowski space-time plot and its
 consequences}  
\vspace*{1.5cm}
\end{center}
\begin{center}
{\bf J.H.Field }
\end{center}
\begin{center}
{ 
D\'{e}partement de Physique Nucl\'{e}aire et Corpusculaire
 Universit\'{e} de Gen\`{e}ve . 24, quai Ernest-Ansermet
 CH-1211 Gen\`{e}ve 4.
}
\newline
\newline
   E-mail: john.field@cern.ch
\end{center}
\vspace*{2cm}
\begin{abstract}
   A sign error in an angle while drawing the original Minkowski plot has persisted
   for a century in text books and the pedagogical literature. When it is corrected,
   the `length contraction' effect derived from the geometry of the plot disappears.
    It is also shown how the `relativity of simultaneity' effect that has been
    derived from the plot results from a lack of correspondence between certain
    geometrical projections on the plot and the properties of
    the physical system ---two spatially
    separated and synchronised clocks in a common inertial frame--- that they are
     purported to describe.

\vspace*{1cm}
\end{abstract}
\end{titlepage}
 
\SECTION{\bf{Introduction}}
    Errors, originating in Einstein's 1905 special relativity
    paper~\cite{Ein1}, in the standard text book interpretation of the physics of the space-time
    Lorentz transformation (LT), have been pointed out in a series of recent papers by the present
    author~\cite{JHFSSC,JHFCRCS,JHFLTASC}. In these papers it is shown that the `length contraction'
    and `relativity of simultaneity' effects, derived directly from the LT, are spurious,
    resulting from a failure to include important additive constants in the equations.
    Einstein pointed out the necessity to include such constants, to correctly describe
    synchronised clocks, in Ref.~\cite{Ein1}, but never actually did so himself. Since the
    demonstration of the spurious nature of the effects is simple, straightforward
    and brief, it is recalled, for the reader's convenience, in the following section
    of the present paper.
    \par The remainder of this paper is devoted to a discussion of the physics of
     the Minkowski space-time plot. In his original `space-time' 
     paper~\cite{Mink} Minkowski 
     derived a `length contraction' effect from the geometry of the plot without
     considering directly the LT. Similar derivations are to be found in many text books
    on special relativity or classical electromagnetism. In Section 3, the projective-geometrical
    properties of the space-time plot are derived from the LT. In Section 4, Minkowski's
    original derivation of `length contraction' is reviewed, and shown to result
    from an erroneous assumption concerning the direction of the world line
    of the considered, uniformly moving, object. In fact, the world line corresponding
    to Minkowski's choice of $x'$ and $t'$ axes is $x = - vt$, whereas it is assumed 
     to be $x = vt$ and therefore to lie along the $t'$ axis. The same mistake is made 
     in all (with, to the present writer's best knowledge, a single exception)
      text-book treatments of the problem.
      Some examples are discussed in Section 5. Also discussed in Section 5 is the
      fortuitously correct derivation of time dilatation from the standard,
      incorrect, Minkowski
      plot, as well as the illusory nature of the `relativity of simultaneity'
      effect suggested by superficial inspection of the plot. This error is not,
      as in the case of `length contraction', the result of a trivial geometrical
      mistake, but arises from a naive interpretation of purely mathematical projection
      operations on the plot that are unrelated to the basic physics of the problem ---observation in
      different inertial frames of events of two spatially-separated and synchronised
      clocks. Section 6 contains a brief summary.  

\SECTION{\bf{The physics of the space-time Lorentz transformation: time dilatation
         and invariant lengths}}
       The LT relates space and time coordinates as measured in two inertial frames
      in relative motion. In the following, it is assumed that the frame S' moves
      with velocity $v$ relative to S, along the direction of a common $x$-$x'$axis.
     In particular, the world lines of the origin, O', of S' with spatial
      coordinates $x'$,$y'$,$z'$=0,0,0 and the fixed point, P',
      with  $x'$,$y'$,$z'$ = L',0,0 are considered. The LT relating the space-time 
    coordinates of O' in S and S' is:
        \begin{eqnarray}
       x'(\OP) & = & \gamma[x(\OP)-v t] = 0 \\
       t'(\OP) & = & \gamma[t-\frac{v x(\OP)}{c^2}]
        \end{eqnarray}
       where $\beta \equiv v/c$, $\gamma \equiv 1/\sqrt{1-\beta^2}$. 
      The space transformation equation (2.1) is equivalent to:
       \begin{equation}
      x'(\OP) = 0 
     \end{equation}
     giving the fixed position of O' in S', and
    \begin{equation}
      x(\OP) = vt
   \end{equation}
    which is the equation of motion (or world line) of O' in S.
    The LT relating the space and time coordinates of P' in S and S' is:
          \begin{eqnarray}
       x'(\PP)-\LP & = & \gamma[x(\PP)- \LA-v t] = 0 \\
       t'(\PP) & = & \gamma[t-\frac{v (x(\PP)- \LA)}{c^2}]
        \end{eqnarray} 
     where (2.5) is equivalent to:
      \begin{equation}
          x'(\PP) = \LP 
      \end{equation}
  and 
       \begin{equation}
          x(\PP) = vt + \LA
       \end{equation} 
        giving, respectively, the position of P' in  S' and its
        equation of motion in S.
      In these equations $t$ is the time recorded by a clock at rest at an
     arbitary position in S~\footnote{As usual, an array of synchronised clocks
     in S may be introduced so that the comparison of $t$ with $t'(\OP)$ and $t'(\PP)$
     can be performed locally.} while $t'(\OP)$ and $t'(\PP)$ are the times recorded by
     similar clocks at O'and L' in S', as observed in S.
     The clocks are set so that when $t = 0$, then $x(\OP) = 0$,
     $x(\PP) = \LA$ and $t'(\OP) =  t'(\PP) = 0$, so the clocks in S' are synchronised at this instant.
     (2.1) and (2.2) are recovered from (2.5) and (2.6) when  $\LA = \LP = 0$.
      It follows from (2.8) that:
      \begin{equation}
      \LA = \left. x(\PP) \right|_{t = 0}
       \end{equation}
       L is therefore a constant that is independent of $v$, defined only by the choice of coordinate
       origin in S.
      As $v \rightarrow 0$, $\gamma \rightarrow 1$, S $\rightarrow$ S' and  $x \rightarrow x'$, so that for $v = 0$, (2.5) is written:
         \begin{equation}
   x'(\PP)-\LP =  x'(\PP)- \LA
     \end{equation}
          so that 
        \begin{equation}
          \LP = \LA
     \end{equation}
     It then follows from (2.3),(2.4),(2.7) and (2.8) that:
          \begin{equation}
   x'(\PP)- x'(\OP) = x(\PP)- x(\OP) =  \LA 
 \end{equation}
      The spatial separation of P' and O' is therefore the same in S' and S for all values of $ x(\PP)$ and $t$
        ---there is no `relativistic length contraction' effect.
     \par   Using (2.4) to eliminate $x(\OP)$ from (2.2), and (2.8) to eliminate $x(\PP)$ from (2.6) gives
     the Time Dilatation (TD) relations:
       \begin {equation}
       t = \gamma t'(\OP) =  \gamma t'(\PP)
       \end{equation}
      The clocks at O' and P' therefore remain synchronised at all times: $t'(\OP) =  t'(\PP)$
       ---there is no `relativity of simultaneity' effect. 
    \par The spurious `length contraction' effect as derived from the LT
   in Einstein's 1905 special relativity paper~\cite{Ein1}, and the associated 
  `relativity of simultaneity' effect are the consequence of using, in the present problem,
    an incorrect LT to describe the world line of P'. Instead of (2.5) and (2.6), the LT (2.1)
     and (2.2), appropriate for O', is used also for P', thus neglecting the additive constants that 
      must be, according to Einstein~\cite{Ein2}, added to the right sides of the latter
       equations in order to correctly describe a synchronised clock at $x' = {\rm L}$ . This gives:
           \begin{eqnarray}
       x'(\PP) & = & \gamma[x(\PP)-v t] =  \LP \\
       t'(\PP) & = & \gamma[t-\frac{v x(\PP)}{c^2}]
        \end{eqnarray}
   Combining (2.14) and (2.1) gives, at any instant in S:
   \begin{equation}
     x'(\PP)- x'(\OP) =  \LP =  \gamma[x(\PP)-x(\OP)] =  \gamma {\rm L}
   \end{equation} 
     This is the spurious `length contraction' effect.
     A universal sign error in drawing the axes in the Minkowski plot has resulted, as shown below,
 in the same false prediction. 
        Combining (2.15) with (2.2), in which $t(\OP) =  t(\PP) = t$, gives
   \begin{equation} 
  t'(\PP)- t'(\OP) = -\frac{\gamma v x(\PP)}{c^2}
    \end{equation} 
     Events which are simultaneous in S are then not so in S'. This is the
       `relativity of simultaneity' effect. How the same spurious effect results from
      misinterpretation of the Minkowski space-time plot is explained below.
    
\SECTION{\bf{Projective geometry of the space-time Lorentz transformation}}
   It is assumed that the LT is used to describe space and time measurements of two objects, 
   O1 and O2, that are at rest along the $x'$ axis of the frame S', separated by a distance $\LP$,
   O1 being at the origin of S' and O2 at $x' = \LP$. As above, the frame S' moves with uniform velocity
   $v$ along the positive $x$-axis of the frame S, the $x$ and $x'$ axes being parallel.
   Space and time measurements of O1 in the frames S and S' are related by the LT:
   \begin{eqnarray}
     x' & = & \gamma(x-\beta x_0) = 0 \\
    x'_0 & = & \gamma(x_0-\beta x)
    \end{eqnarray}
    where $x_0 \equiv ct$ and $x'_0 \equiv ct'$.
    These equations may be written as two-dimensional rotations by introducing the variables:
   \begin{eqnarray}
   \cos \theta & \equiv & \frac{1}{\sqrt{1 + \beta^2}} \\
   \sin \theta & \equiv & \frac{\beta}{\sqrt{1 + \beta^2}} \\
    X' & \equiv & x'\sqrt{\frac{1 -\beta^2}{1 +\beta^2}} \\
    X'_0 & \equiv & x'_0 \sqrt{\frac{1 -\beta^2}{1 +\beta^2}}
   \end{eqnarray}
    as
  \begin{eqnarray}
   X' & = & x \cos \theta- x_0 \sin \theta  \\
  X'_0 & = & x_0 \cos \theta- x \sin \theta
  \end{eqnarray}
   These equations show that the LT of both space and time coordinates may be written
   as the product of a two-dimensional rotation and a scale transformation, the scale
   factor being $\sqrt{1 +\beta^2}/\sqrt{1-\beta^2}$ for both coordinates. It is easily
   shown that the $X'$-axis is obtained from the  $x$-axis by clockwise rotation 
   through the angle $\theta = \arctan \beta$, while the  $X'_0$-axis is obtained from the
   $x_0$-axis by anti-clockwise rotation through the same angle. Fig.1 shows the
   $x,x_0 \rightarrow X'_0$ transformation of Eqn(3.8).

\begin{figure}[htbp]
\begin{center}\hspace*{-0.5cm}\mbox{
\epsfysize8.0cm\epsffile{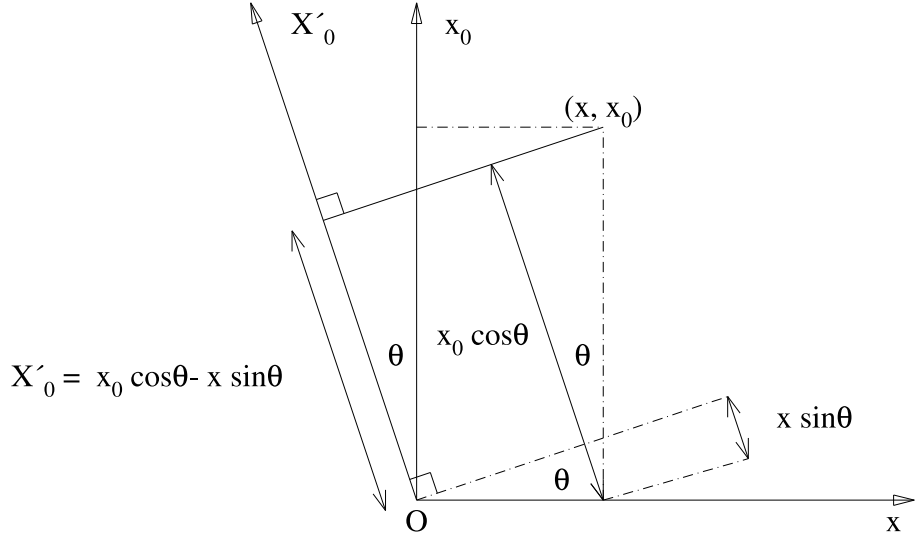}}
\caption{ {\em  The Lorentz transformation from $x$,$x_0$ to $X'_0$ using Eqn(3.8).}}  
\label{fig-fig1}
\end{center}
\end{figure}

   \par In Fig.2 is shown the world line WL1 of O1: $\Delta x = \beta \Delta x_0$ on a 
    Minkowski plot where $x$ and $x_0$ are described by rectangular Cartesian coordinates. The
     corresponding $X'$ , $X'_0$ axes are also drawn. It can be seen that the projection
     needed to obtain $\Delta x'_0$ from $\Delta x$ and $ \Delta x_0$ is neither
 orthogonal (perpendicular to the  $X'_0$ axis) nor oblique (parallel to the  $X'$ axis).
  The geometry of Fig.2 and the relation (3.6) gives, for the projection angle, $\phi_t$:
    \begin{equation}
   \tan \phi_t = \frac{2 \beta}{\sqrt{1-\beta^2}(\sqrt{1+\beta^2}-\sqrt{1-\beta^2})}
    \end{equation}
   This angle takes the limiting value $\pi/2$ in both the
   $\beta \rightarrow 0$ and  $\beta \rightarrow 1$ limits. Differentiation
   of (3.9) shows that  $\phi_t$ takes its minimum value when $\beta = \beta_{min}$
    where:
   \begin{equation}
    (\beta_{min})^6+(\beta_{min})^4+(\beta_{min})^2-1 = 0
    \end{equation}
   Solving this cubic equation for $(\beta_{min})^2$ gives $\beta_{min} = 0.737$ corresponding to
    $\phi_t^{min} = 75.44^{\circ}$.
    \par From the geometry of Fig.2 and Eqn(3.6), 
    \begin{eqnarray}
       \Delta x'_0 & = & \sqrt{\frac{1 +\beta^2}{1 -\beta^2}} \Delta X'_0  = 
    \sqrt{\frac{1 +\beta^2}{1 -\beta^2}}\left[ \sqrt{(\Delta x)^2 +(\Delta x_0)^2}\right]
     \cos 2 \theta \nonumber \\
       & = & \frac{1+\beta^2}{\sqrt{1 -\beta^2}} \Delta x_0 (\cos^2\theta- \sin^2\theta) =
         \frac{ \Delta x_0(1-\beta^2)}{\sqrt{1 -\beta^2}}  \nonumber \\
       & = & \frac{\Delta x_0}{\gamma} 
   \end{eqnarray}
     so that $\Delta x_0 = \gamma  \Delta x'_0$, the well-known and experimentally-verified
        time dilatation (TD) effect.

\begin{figure}[htbp]
\begin{center}\hspace*{-0.5cm}\mbox{
\epsfysize12.0cm\epsffile{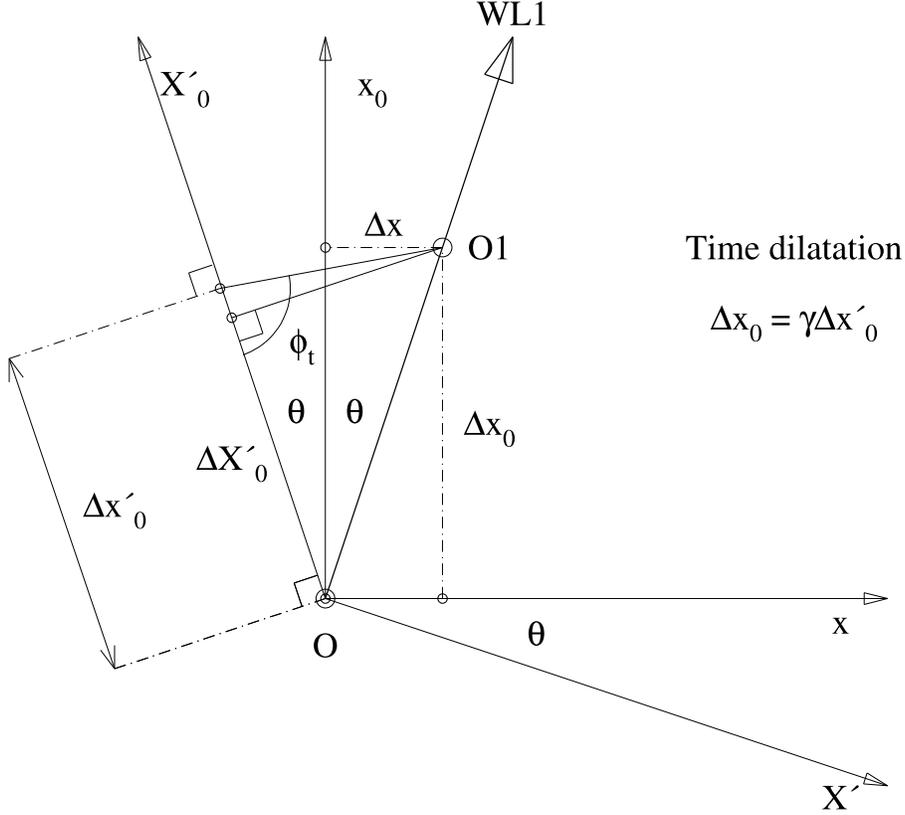}}
\caption{ {\em  Transformation of the intervals $\Delta x$ and $\Delta x_0$ on the world line
      $\Delta x = \beta \Delta x_0 = \Delta x_0\tan \theta$ in the frame S into
    the frame S' according to Eqns(3.6)-(3.8). In this and subsequent figures large circles 
    denote world-line points and small circles projected coordinates. The LT is equivalent
  to an orthogonal projection onto the $X'_0$ axis multiplied by the scale factor of
  Eqn(3.6). This gives the time dilatation relation (3.11).}}  
\label{fig-fig2}
\end{center}
\end{figure}

    \par  The  space-time LTs for O1 and O2
     may be written (c.f. Eqns(2.1),(2.2),(2.5),(2.6) and (2.11)) as:
  \begin{eqnarray}
          x'({\rm O}1)& = & \gamma[x({\rm O}1)-\beta x_0] = 0 \\
    x'_0& = & \gamma[x_0-\beta x({\rm O}1)] \\
        x'({\rm O}2)-{\rm L}  & = & \gamma[x({\rm O}2)-{\rm L} -\beta x_0] = 0 \\
    x'_0 & = & \gamma[x_0-\beta(x({\rm O}2)-{\rm L})]
  \end{eqnarray}
    where $x_0$ and  $x'_0$ correspond to the times recorded by any clock in S and S', respectively,
     synchronised so that $x_0 = \gamma x'_0$.
   Defining  $x1 \equiv  x({\rm O}1)$, $x1' \equiv  x'({\rm O}1)$,
    $x1_0 = x2_0 \equiv x_0$, $x1'_0 = x2'_0 \equiv x'_0$, and
   introducing for O2 the coordinate transformations: $x2 \equiv x({\rm O}2)-{\rm L}$,
   $x2' \equiv x'({\rm O}2)-{\rm L}$, (3.12),(3.13) and (3.14),(3.15) are written, in a similar manner, as:
    \begin{eqnarray}
    x1' & = & \gamma(x1-\beta x1_0) = 0 \\
  x1'_0 & = & \gamma(x1_0-\beta x1) \\
x2' & = & \gamma(x2-\beta x2_0) = 0 \\
  x2'_0 & = & \gamma(x2_0-\beta x2)
 \end{eqnarray}
   The world lines WL1 and WL2 referred to the coordinate systems, in the frame S,
   ($x1$,$x1_0$) and($x2$,$x2_0$), respectively, with origins at O$_1$ and O$_2$ are shown
    in Fig.3. Along the world lines, $x1=\beta x1_0$ and $x2=\beta x2_0$,
     the synchronisation condition  $x1_0 = x2_0$ is relaxed so that
     $x1_0$ and $x2_0$ are allowed to vary independently in (3.16) and (3.18).
    Also shown are the loci of the positions of O1 and O2 (shown in the plot for
     $\beta = 1/3$) for other values of $\beta$. These are the hyperbolae:
     \begin{eqnarray}
       x1_0^2-x1^2 & = & (c\tau)^2~~~~~ x1,x1_0 \ge 0 \\
       x2_0^2-x2^2 & = & (c\tau)^2~~~~~ x2,x2_0 \ge 0
   \end{eqnarray}
  \par The absence of any length contraction effect is made manifest by the constant 
   separation, $\LA$, of the hyperbolae for any value of $x_0$ and $\beta$, including
    $\beta = 0$, when $x_0 = c \tau$.
     \par  Also clear, by inspection of Fig.3 is the absence
    of any relativity of simultaneity effect. Introducing the variables defined in
    Eqns(3.3)-(3.6) into Eqns(3.16)-(3.19) gives:
 \begin{eqnarray}
   X1' & = & x1 \cos \theta- x1_0 \sin \theta  \\
  X1'_0 & = & x1_0 \cos \theta- x1 \sin \theta \\
  X2' & = & x2 \cos \theta- x2_0 \sin \theta  \\
  X2'_0 & = & x2_0 \cos \theta- x2 \sin \theta
  \end{eqnarray}
  The $X1'$, $X1'_0$, $X2'$ and $X2'_0$ axes for $\beta = 1/3$ are drawn in Fig.3.
  Since:
   \begin{equation} 
     {\rm O}_1{\rm N}_1 =  X1'_0 = {\rm O}_2{\rm N}_2 =  X2'_0
   \end{equation}
  when $x1_0 = x2_0$ it follows that, at that instant, $x1'_0 =  x2'_0$ so that spatially
  separated events that are simultaneous in S (events on the world lines of
  O1 and O2 when  $x1_0 = x2_0 = x_0$) are also simultaneous in S' --- $x1'_0 =  x2'_0 = x'_0$.

\begin{figure}[htbp]
\begin{center}\hspace*{-0.5cm}\mbox{
\epsfysize15.0cm\epsffile{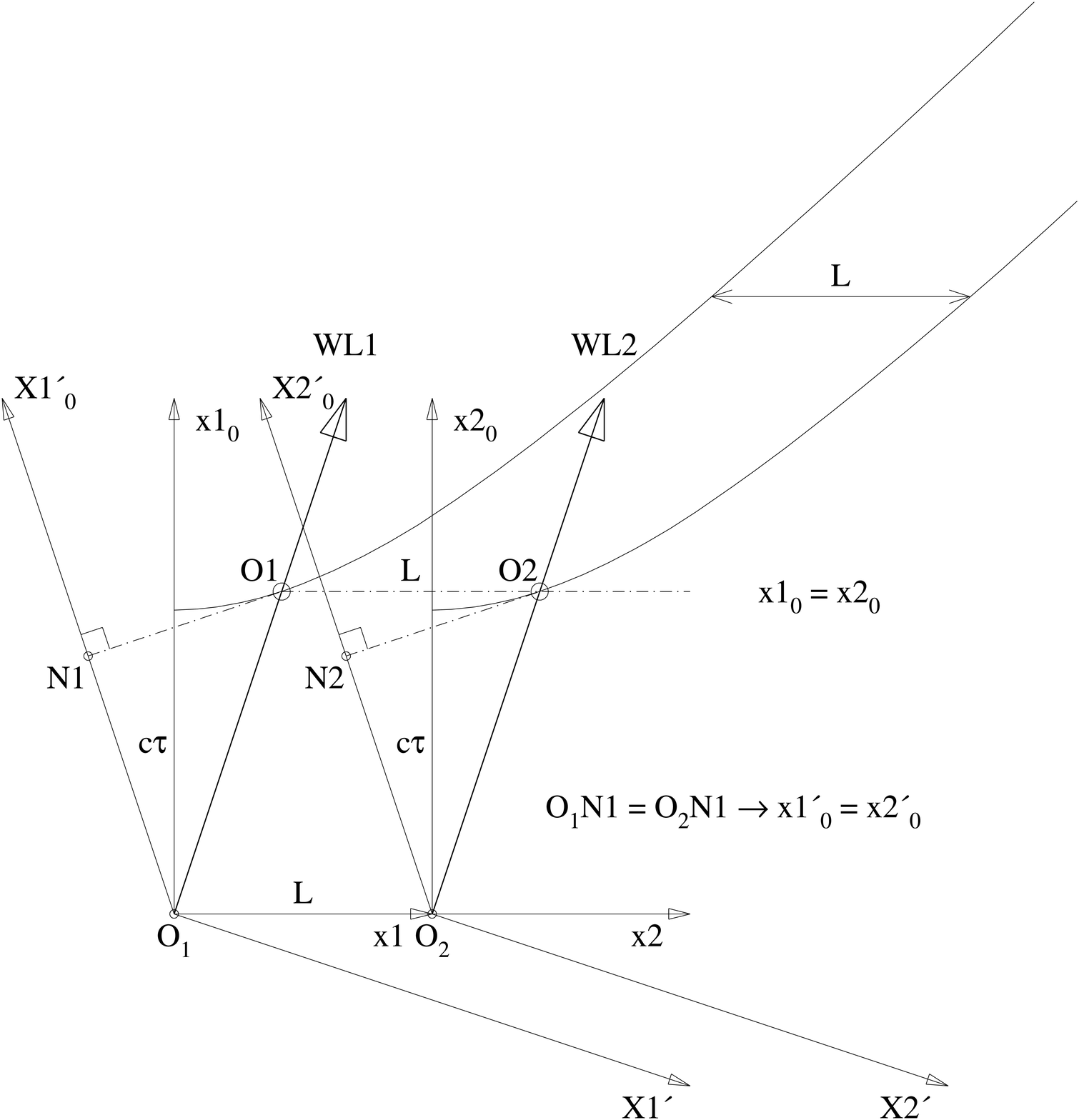}}
\caption{ {\em Projection of simultaneous events in S on the world lines of 
    O1 and O2 onto the $X1'_0$ and $X2'_0$ axes defined by Eqns(3.23) and (3.25).
     WL1 and WL2 correspond to $\beta = \tan \theta = 1/3$. World line points for O1 and O2
     for other values of $\beta$, when $x1_0 = x2_0 = c\tau$ for $\beta = 0$, lie on the hyperbolic curves passing 
     through the indicated positions of  O1 and O2. 
    The absence of any `relativity of simultaneity' or `length contraction' effects
   is evident from inspection of this figure, which shows invariance under the
    transformation of spatial coordinates $x \rightarrow x +\LA$, for O1 $\rightarrow$ O2.}}  
\label{fig-fig3}
\end{center}
\end{figure}

  \par Indeed, simple inspection of Fig.3 shows that that the absence of any length contraction
   or relativity of simultaneity effect is a necessary consequence of translational
   invariance ---the world line WL2 being obtained from WL1 by the spatial coordinate 
   substitution $x \rightarrow x +\LA$ (compare Eqns(3.12) and (3.13) with (3.14) and (3.15)).
   \par In Fig.4 are shown world lines and coordinate axes for objects at rest in S'
     for different values of $\beta$, as well as the hyperbolic loci of the points
      shown on WL($\beta=1/3$) and WL($\beta=1/2$) for other values of $\beta$. At large
    values of $x_0$ the hyperbola approaches its asymptote, the projection into
    the $x$, $x_0$ plane of the light cone which is the world line WL($\beta=1$).
     It is important to note that for a given orientation of transformed axes: 
      $X'$,$X'_0$; $X''$,$X''_0$:.. the value of $\beta$ and hence the slope of the corresponding
      world line in S is fixed. The  $X'$,$X'_0$ axes correspond to $\beta = 1/3$ and
        the  $X''$,$X''_0$ axes to $\beta = 1/2$. It is nonsensical to draw on the plots world lines
      with any
      other slope. As will be seen in the following section, just this fundamental error
      was made by Minkowski in his original `space-time' paper~\cite{Mink}, and has since been
      universally followed by authors of text books and the pedagogical literature.

   \begin{figure}[htbp]
\begin{center}\hspace*{-0.5cm}\mbox{
\epsfysize15.0cm\epsffile{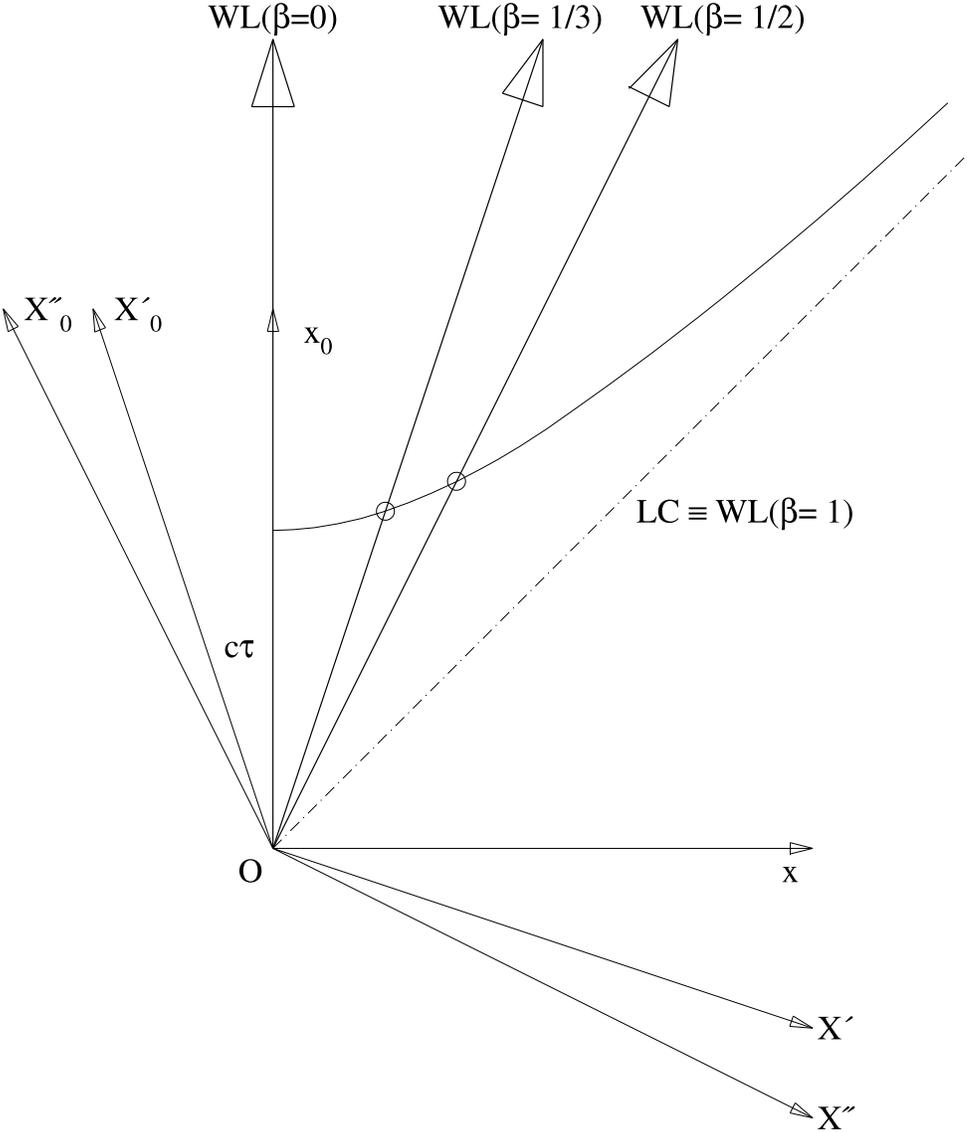}}
\caption{ {\em  Coordinate axes in S' for different world lines in S.
    There is a one-to-one correspondence between world line directions and the orientation
   of the axes in S': $X'$,$X'_0$ correspond to $\beta = 1/3$ and
    $X''$,$X''_0$ to $\beta = 1/2$. Drawing any other world lines for the S' coordinate
     axes shown is a nonsensical procedure.}}  
\label{fig-fig4}
\end{center}
\end{figure}

\SECTION{\bf{Minkowski's spurious `length contraction' effect}}
   A fair copy of Fig.1 of Ref.~\cite{Mink}, used by Minkowski to derive a `length contraction'
   effect, is drawn in Fig.5. Only Minkowski's time coordinates $t$,$t'$ are replaced by
   $x_0$,$x'_0$ and, for clarity, the intersection of the line A'B' ---the tangent to the
    hyperbola AO$^2 = x_0^2-x^2$ at A'--- with the $x_0$ axis, has been labelled E'.
     Minkowski assumes that A', a point on the world line of the orgin of S', lies on the
     $x'_0$ axis, for non-zero values of $x'_0$ . This is wrong.
      If the world line of the origin of S', as viewed in S, 
     is $x = \beta x_0$. as assumed by Minkowski, the  $x'$, $x'_0$ axes must be drawn parallel
    to the $X'$, $X'_0$ axes shown in Fig.4 above. The correct world line for the  $x'$, $x'_0$
     axes of Fig.5a is  $x = -\beta x_0$, as shown in Fig.6. In fact the hyperbola giving the
     locus of the coordinates of the origin of S', for different values of $\beta$, can never 
     cross the $ x'_0$ axis, as shown in Fig.5a. The LT is not mentioned in Minkowski's
     paper ---only geometric properties of the hyperbolae which specify time-like or
     space-like invariant intervals. To calculate coordinates from invariant interval
     relations square roots must be taken, leading to sign ambiguities. Insufficient
     attention to the correct sign choice in correlating world lines with axis directions
     is at the origin of Minkowski's error.
     \par In order to calculate a `length contraction' effect Minkowski assumed an oblique
     projection as shown in Fig.5b. Thus the length Q'Q' of an object lying along the 
      $x'$ axis corresponds to the length QQ along the $x$-axis, the lines QQ' being
       drawn parallel to the $x'_0$ axis. Such a projection is seen to result in a length
       contraction effect of QQ as compared to Q'Q'. However, the geometry of Fig.5b gives:
      \begin{equation}
       {\rm QQ} = \frac{{\rm Q'Q'}(1-\beta^2)}{\sqrt{1+\beta^2}}
      \end{equation}
       not the conventional length contraction effect
       which would instead give: ${\rm QQ} ={\rm Q'Q'}\sqrt{1-\beta^2}$. In order to arrive at the latter result,
       Minkowski makes the following hypotheses connecting the geometries of Fig.5a and Fig.5b:  
         \begin{eqnarray}
        {\rm PP} & = & \ell  {\rm OC} \\
        {\rm Q'Q'} & = & \ell  {\rm OC'} \\
        {\rm QQ} & = & \ell  {\rm OD'}
     \end{eqnarray} 
      where $\ell$ is the length along the $x'$ axis of an object at rest in S' and
     observed in S', and uses a different geometrical definition of the length contraction
     effect. Minkowski's arguments are based on the geometry of Fig.5b.
      However the axis O$x'$ in the figure is inconsistent with the world lines PP' of the ends of
     an object at rest in S. For such an object $\beta = 0$ and the $x$ and $x'$ axes coincide.  
      Taking the ratio of (4.4) to (4.2) gives:

      \begin{equation}
      \frac{{\rm QQ}}{{\rm PP}} =  \frac{{\rm OD'}}{{\rm OC}} \equiv f_{LC}
 \end{equation}
      Minkowski now assumes that $f_{LC}$ is the length contraction factor. Since the line 
          E'A'B' is drawn as a tangent to the hyperbola, it follows that the tangent
          of the angle AE'A' is $1/\beta$ and the line segment E'A'B' is parallel to the $x'$ axis. From the
          symmetry of the figure about the light cone projection OBB', the angles AE'A'
          and CD'C' are equal. It then follows from the geometry
          of Fig.5a that ${\rm OD'} ={\rm OC}\sqrt{1-\beta^2}$ giving, with (4.5), the relation
        \begin{equation}
         {\rm QQ} ={\rm PP}\sqrt{1-\beta^2} 
 \end{equation}
  or $ f_{LC} = \sqrt{1-\beta^2}$.
 
   \begin{figure}[htbp]
\begin{center}\hspace*{-0.5cm}\mbox{
\epsfysize15.0cm\epsffile{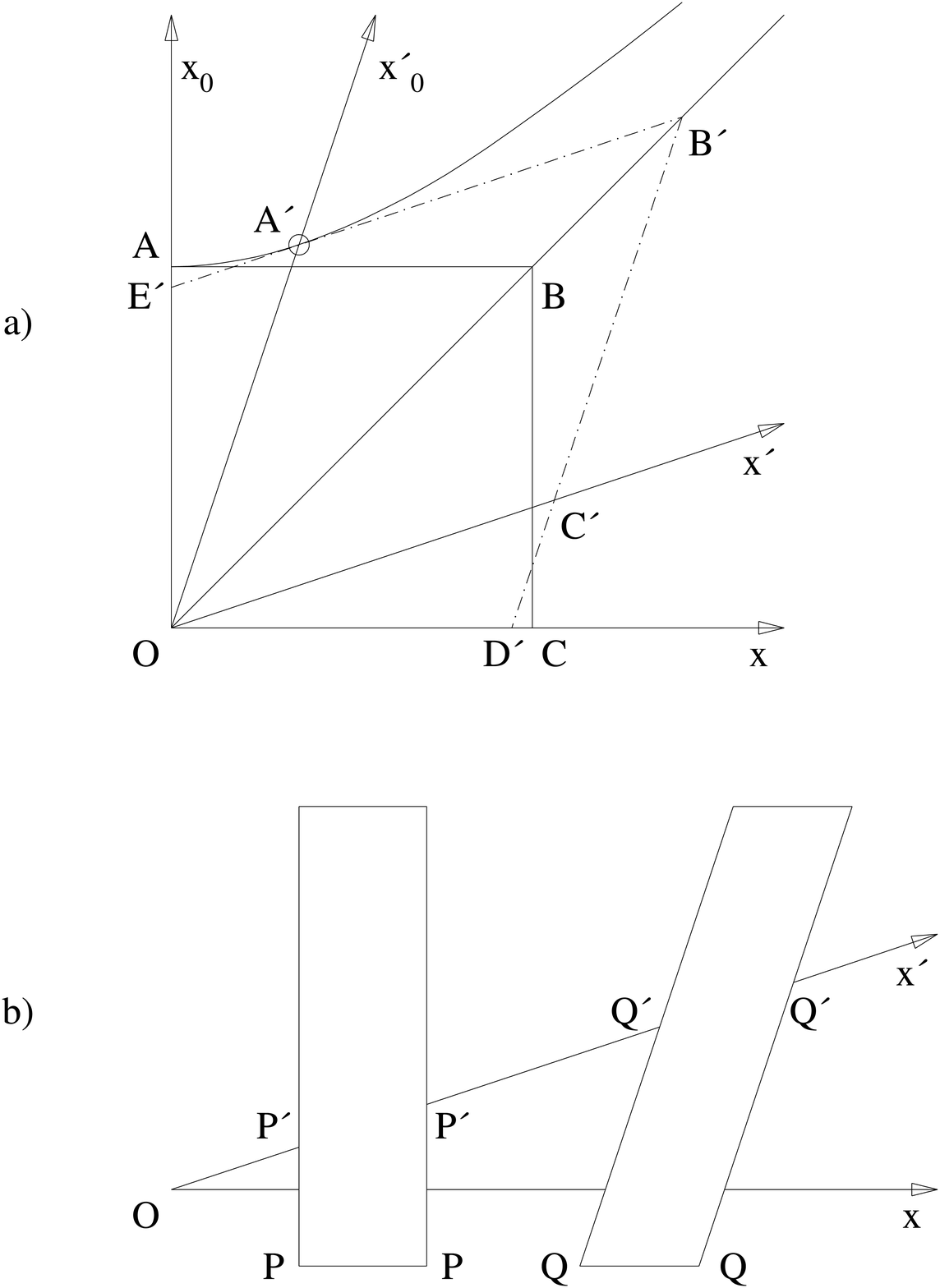}}
\caption{ {\em Copy of Minkowski's figure from Ref.~\cite{Mink}. See text for
    discussion.}}  
\label{fig-fig5}
\end{center}
\end{figure} 

      The argument given attempts to relate the world lines of the ends an object at rest in S,
     represented
      by  PP', to the world lines, QQ', of the ends of an object at rest in S',
      on the assumption that the $x'$ axis is as drawn
      in Fig.5a. This is a incorrect procedure since the axes needed
       to specify the world line, in the frame S', of an object at rest in the frame S, are not as
       shown in Fig.5b, but as in Fig.6 with exchange of primed and unprimed coordinates. This
       space-time plot is discussed further in the following section (see Fig.13c below).   
       The fundamental error in Fig.5a of incorrectly plotting the world line of the origin
       of S' along the $x'_0$ axis, instead of as in Fig.6, has been systematically repeated in almost
       all text book discussions of the Minkowski plot. It will be seen however that the argument
       typically used to obtain the length contraction effect, although, finally, geometrically 
      identical
       to Minkowski's calculation, does not invoke the world lines PP' of an 
         object at rest in S, but considers,
        instead of the time-like interval relation invoked by Minkowski
        (the hyperbola through A and A'), a space-like one involving directly the length of
         the considered object.

\begin{figure}[htbp]
\begin{center}\hspace*{-0.5cm}\mbox{
\epsfysize8.0cm\epsffile{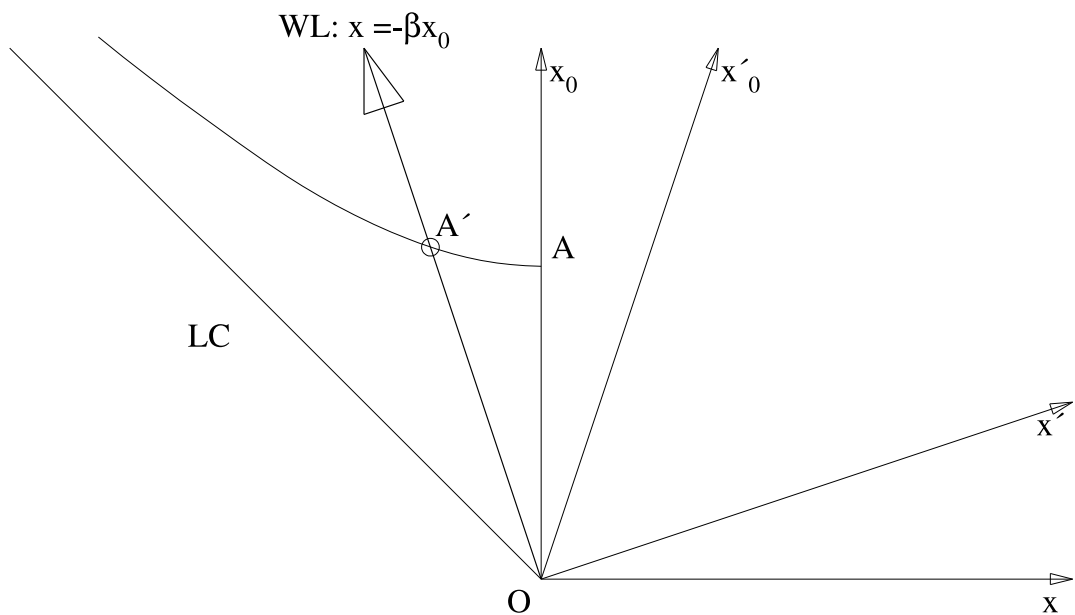}}
\caption{ {\em Correct world line in S of the origin of S' for Minkowski's choice
   of $x$ and $x'_0$ axes as in Fig.5a. Compare with the incorrect world line OA' in 
   the latter figure.}}  
\label{fig-fig6}
\end{center}
\end{figure}

\SECTION{\bf{Text book treatments of the Minkowski plot}}
    For definiteness, the discussion of the Minkowski plot in the widely-known book `Spacetime Physics'
     by Taylor and Wheeler ~\cite{TW}
     will be first considered. Similar treatments are to be found in the books of Aharoni~\cite{Aharoni},
    Panofsky and Philips~\cite{PP} and Mermin~\cite{Mermin}. Minkowski's sign error in drawing the directions of the $x'$ 
     and $t'$ axes has been replicated in many text books treating special relativity. A survey
      of such books in the library of the `Section de Physique' at the University
      of Geneva and the CERN library, turned up 24 books with the error, including notably
     the first volume of `The Feynman Lectures in Physics'~\cite{FeynLPV1}. The only exception
     found was a book by Anderson~\cite{Anderson} where the world line of S' 
      corresponding to the $x'$, $x'_0$ axes as drawn in Fig.5 is correctly given as $x = -\beta x_0$
     corresponding to the LT: $x' = \gamma(x +\beta x_0) = 0$. However, this author does not
    use the Minkowski plot to discuss time dilatation or `length contraction', but rather the
     analysis of light signals observed in different inertial frames.
     \par Fig.7, which is a combination of Figs.66 and 69
     of Ref.~\cite{TW} is the basis for discussion of both time dilatation and `length contraction'
     in this book.   
     An object at rest in S', lying along the $x'$ axis, is considered and the world lines
     of the ends $\sharp$ 1 and $\sharp$ 2 of the object are drawn, end $\sharp$ 1 being at the origin of S'. As in    the 
     original Minkowski plot, Fig.5a, the world line of the origin in S' is incorrectly assumed to
     lie along the $x'_0$ axis instead of on the other side of the $x_0$ axis, as in Fig.6.

\begin{figure}[htbp]
\begin{center}\hspace*{-0.5cm}\mbox{
\epsfysize8.0cm\epsffile{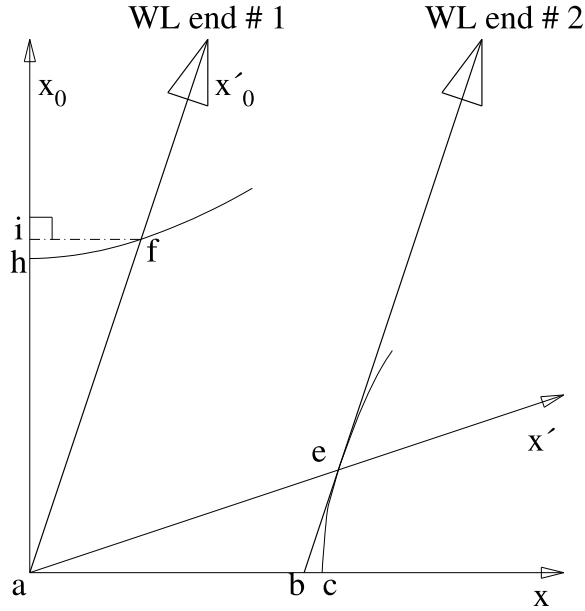}}
\caption{ {\em  Figures used in Ref.~\cite{TW} to discuss the time dilatation and `length contraction'
     effects. See text for discussion.}}  
\label{fig-fig7}
\end{center}
\end{figure}

     \par In order to calculate the time dilatation effect it is assumed that the distance $af$ in 
      the frame S represents one unit of time in the frame S', and that the distance $ah$ represents
      one unit of time in the frame S. Since $h$ and $f$ lie on the hyperbola:
      \begin{equation}
         (ah)^2 = (\Delta x_0)^2-(\Delta x)^2 = (ai)^2-(if)^2 = (ai)^2(1-\tan^2 \theta)
            = (ai)^2(1-\beta^2)
      \end{equation} 
     it follows that 
     \begin{equation}
     \frac{af}{ah} = \frac{ai}{ah \cos \theta}= \sqrt{\frac{1+\beta^2}{1-\beta^2}} \equiv F
    \end{equation}
      so that one unit of time in the frame S' is assumed to correspond to $F$ units of time
       along the $x'_0$ axis in the frame S~\cite{Mermin1}. To derive the time dilatation effect it is further
       assumed that the time in S of the event at $f$ in S' is given by the orthogonal
       projection $fi$ onto the $x_0$ axis. The time dilatation factor is then found to
       be, using (5.2) and (3.3):
    \begin{equation}
       \frac{\Delta x_0}{\Delta x'_0} = \frac{ai}{ah} =\frac{af \cos \theta}{ah}
        = \frac{1}{\sqrt{1-\beta^2}} = \gamma
  \end{equation}
     For direct comparison with the calculation of Section 2 above, using the correct projection
    procedure in the $x$, $x_0$ plane, it is convenient to introduce explicitly into the plot
    the scale factor of Eqn(5.2) relating a coordinate $\tilde{X}'_0$ to $x'_0$:
   \begin{equation}
\tilde{X}'_0 =  \sqrt{\frac{1+\beta^2}{1-\beta^2}} x'_0 \equiv F  x'_0
  \end{equation}
    The scale factor relating $\tilde{X}'_0$ to $x'_0$ is the reciprocal of that
    relating $X'_0$ to $x'_0$ in Eqn(3.6). The calculation of the time dilatation relation
     (5.3), using the orthogonal projection onto the $x_0$ axis, is shown in Fig. 8a, where
      WL1 denotes the world line of end $\sharp$ 1 of the object. The
      geometry of this figure shows that:
   \begin{equation}
        \Delta x'_0 = \frac{\Delta\tilde{X}'_0}{F} = \frac{\Delta x_0}{ F \cos \theta}
            =  \frac{\Delta x_0}{ \gamma}
 \end{equation}
       in agreement with (5.3). The correct calculation of the time dilatation effect
      with the world line of the end $\sharp$ 1 of the object on the opposite side of the $x_0$ axis,
     as in Fig.6, is shown in Fig. 8b. The orthogonal projection is now from the event
     onto the $X'_0$ axis which is scaled up by the factor $F$ to obtain the  $x'_0$
     coordinate according to Eqn(3.6). Since the geometry of Fig.8b is related to that
    of Fig.2 by reflection in the $x_0$ axis, the same result, (3.11), identical to (5.5)
    is obtained. By use of a different projection (orthogonal from the event onto the
    $x_0$ axis, instead of onto the $X'_0$ axis) and a different scaling factor for
     time intervals between S and S' (the reciprocal of that, (3.6), obtained from the LT) the
    correct time dilatation formula is obtained in Fig.8a, from the incorrectly
      drawn world lines of Figs.7 and 8a. As will now be seen, this fortuitous cancellation
     of three different errors is no longer applicable in the discussion of the
     transformation of spatial intervals.

\begin{figure}[htbp]
\begin{center}\hspace*{-0.5cm}\mbox{
\epsfysize17.0cm\epsffile{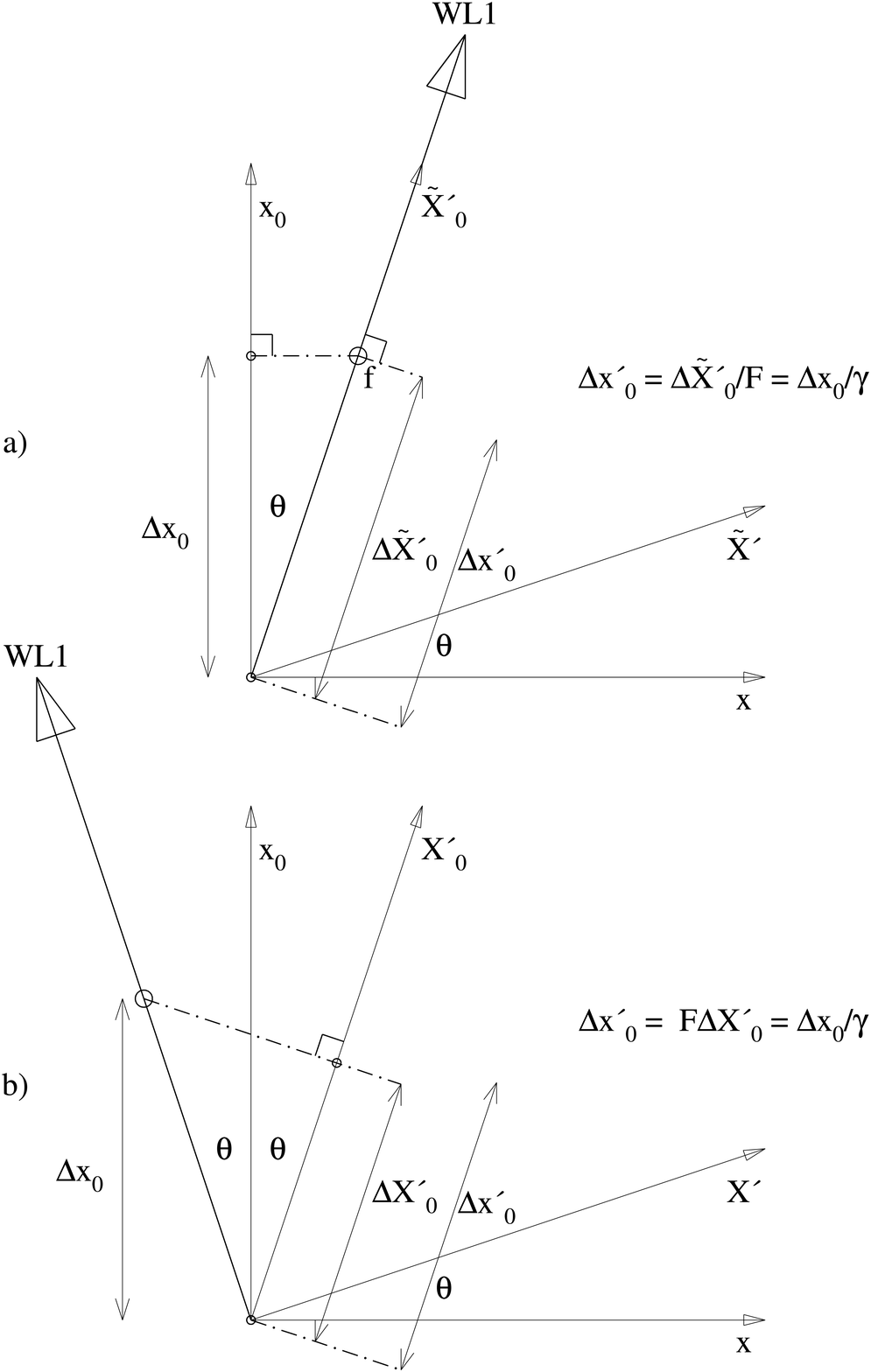}}
\caption{ {\em  a) Calculation of time dilatation according to Ref.~\cite{TW}. b) Correct calculation
    of time dilatation from the LT as in Fig.2 above. See text for discussion.}}  
\label{fig-fig8}
\end{center}
\end{figure}

     \par By considering the hyperbola passing through the points $e$ and $c$ in Fig.7, a relation
      analogous to 5.2 is derived:
     \begin{equation}
     \frac{ae}{ac} = \sqrt{\frac{1+\beta^2}{1-\beta^2}}
    \end{equation}
      If $\Delta x'$ is the distance between $a$ and $e$ in the frame S', it is now assumed, as for
      the discussion of the time coordinates in the time dilatation effect, that
    \begin{equation}
     ae = \sqrt{\frac{1+\beta^2}{1-\beta^2}} \Delta x'
    \end{equation}
     In order to obtain the corresponding length interval in the frame S, an oblique projection,
      parallel to the $x'_0$ axis is taken, instead of the orthogonal one, perpendicular to the $x_0$-axis,
    assumed for the transformation
      of time intervals between S' and S. This gives:
    \begin{equation}
    \Delta x = ab
   \end{equation}
    Combining (5.7) and (5.8) gives:
      \begin{equation}
    \Delta x =  \frac{ab}{ae}\sqrt{\frac{1+\beta^2}{1-\beta^2}}\Delta x' 
   \end{equation}
    Note that this definition of the `length contraction' effect is not the same as 
    that assumed by Minkowski. In terms of the geometrical definitions of Fig.7, Minkowski
    assumes that:
       \begin{equation}
   \Delta x = f_{LC}  \Delta x' = \frac{ab}{ac} \Delta x'
     \end{equation}
   Therefore no scale factor connecting lengths observed in the frame S' to those observed
   in the frame S is introduced by  Minkowski.
   \par It is found from the geometry of Fig.7 that 
      \begin{equation}
       \frac{ab}{ae} = \frac{1-\beta^2}{\sqrt{1+\beta^2}}        
      \end{equation} 
  so that
      \begin{equation}
    \Delta x =  \sqrt{1-\beta^2}\Delta x' = \frac{\Delta x'}{\gamma}
   \end{equation}     
     This is the `length contraction' effect calculated according to Ref.~\cite{TW}. 
     A comparison with a calculation where the world lines of the ends of an 
     object consistent with the $x'$, $t'$ axes drawn in Fig.7 and using the correct
     projection procedure obtained from the LT is shown in Fig.9. For ease of comparison,
     the variable $\tilde{X}'$ analogous to $\tilde{X}'_0$ in Eqn(5.4) is introduced:
    \begin{equation}
\tilde{X}' =  \sqrt{\frac{1+\beta^2}{1-\beta^2}} x' = F  x'
  \end{equation}
     The derivation of Eqn(5.12), as just described, is shown in Fig. 9a
   \par In Fig. 9b the world lines WL1, WL2 of end $\sharp$ 1 and end $\sharp$ 2
    of the object are shown in the frame S for $\beta = 0$ and $\beta = 1/3$. Inspection of Fig. 9b, 
    where the correct projections as derived from the LT, which 
      in the present case are (c.f. Eqns(3.32)-(3.25)):
         \begin{eqnarray}
   X1' & = & x1 \cos \theta+ x1_0 \sin \theta  \\
  X1'_0 & = & x1_0 \cos \theta+ x1 \sin \theta \\
  X2' & = & x2  \cos \theta+ x2_0 \sin \theta  \\
  X2'_0 & = & x2_0 \cos \theta+ x2 \sin \theta
  \end{eqnarray}
   are used, shows that, as in the analogous Fig. 3, the length of the
   object in S' (L') is the same as in S (L) and that there is no `relativity of 
     simultaneity' effect.
    As previously shown in Section 2 above, the length of an object
   (by definition a $\Delta x_0 =  \Delta x'_0 =  0$ projection) is a Lorentz invariant
    quantity --- there is no `length contraction'. For the transformation
   of spatial intervals there is evidently no fortuitous cancellation of the errors of drawing
    wrongly the world lines of the ends of the object, using an oblique projection, and a scaling
    factor that differ from the relations (5.14),(5.15) and (3.5),(3.6) obtained directly from the LT.

 \begin{figure}[htbp]
\begin{center}\hspace*{-0.5cm}\mbox{
\epsfysize23.0cm\epsffile{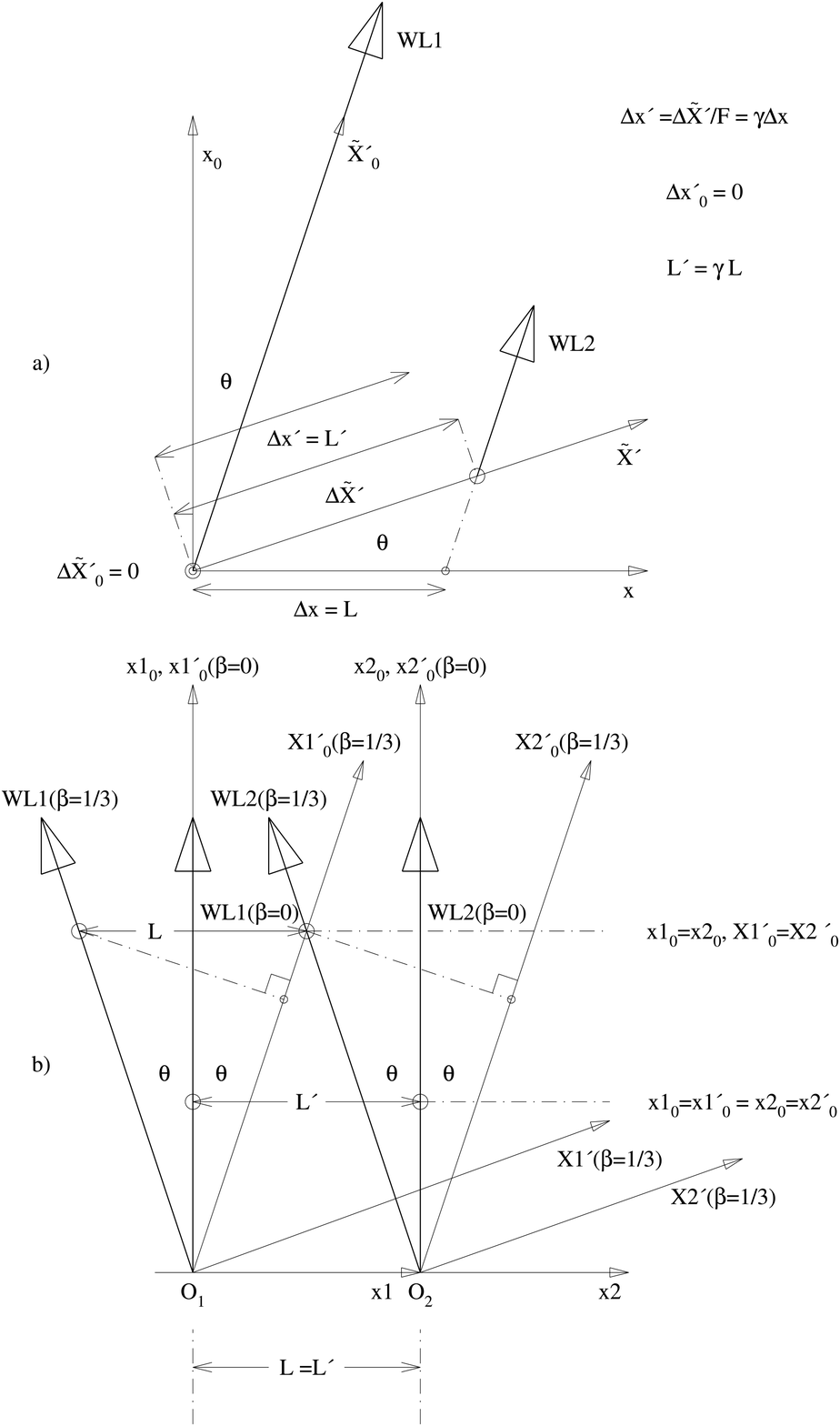}}
\caption{ {\em  a) Calculation of `length contraction' according to Ref.~\cite{TW}. b) Correct prediction
   of the LT showing the invariance of the length of the object and the absence of any
     `relativity of simultanenity' effect. See text for discussion.}}  
\label{fig-fig9}
\end{center}
\end{figure}

     \par Inspection of Fig.7 shows that Ref.\cite{TW} (in common with other text books)
       uses different projections in the discussions of time dilatation and `length contraction'
      ---orthogonal to the $x_0$ axis for time dilatation and oblique, parallel to the $x'_0$
      axis, for `length contraction'. Given the symmetry of the LT equations with respect
      to exchange of $x$,$x_0$ and  $x'$,$x'_0$ also manifest in the correct projection
      equations (3.7) and (3.8) or (5.14) and (5.15), there can be no physical justification
      for the arbitary choice of projections that is made. Using, in each case the other
     type of projection would give time contraction and length dilatation effects, both in
    conflict with the correct predictions of the LT ---time dilatation but no `length
    contraction'--- as well as, in the former case, contradiction to experiment.
     \par This point is
    illustrated in Fig.10 where the transformation of time is calculated using: in a) a projection
    normal to the $x_0$ axis as in Ref.~\cite{TW}, in b) an oblique projection as used in
     Ref.~\cite{TW} to derive `length contraction' and in c) the correct projections
    of the LT using Eqns(3.5)-(3.8), {\it mutatis mutandis} to account for the different
     direction of the world line. Fig.10b corresponds to `time contraction'. Application of
     the perpendicular projection of  Fig.10a  to the $x$-axis would give a 
      `length dilatation' effect. 
     In the correct relativistic analysis of the transformation properties of time
    and length intervals shown in Fig.8b and Fig.9b, respectively, the same projection and
    scaling operations are used in both calculations.

     \begin{figure}[htbp]
\begin{center}\hspace*{-0.5cm}\mbox{
\epsfysize17.0cm\epsffile{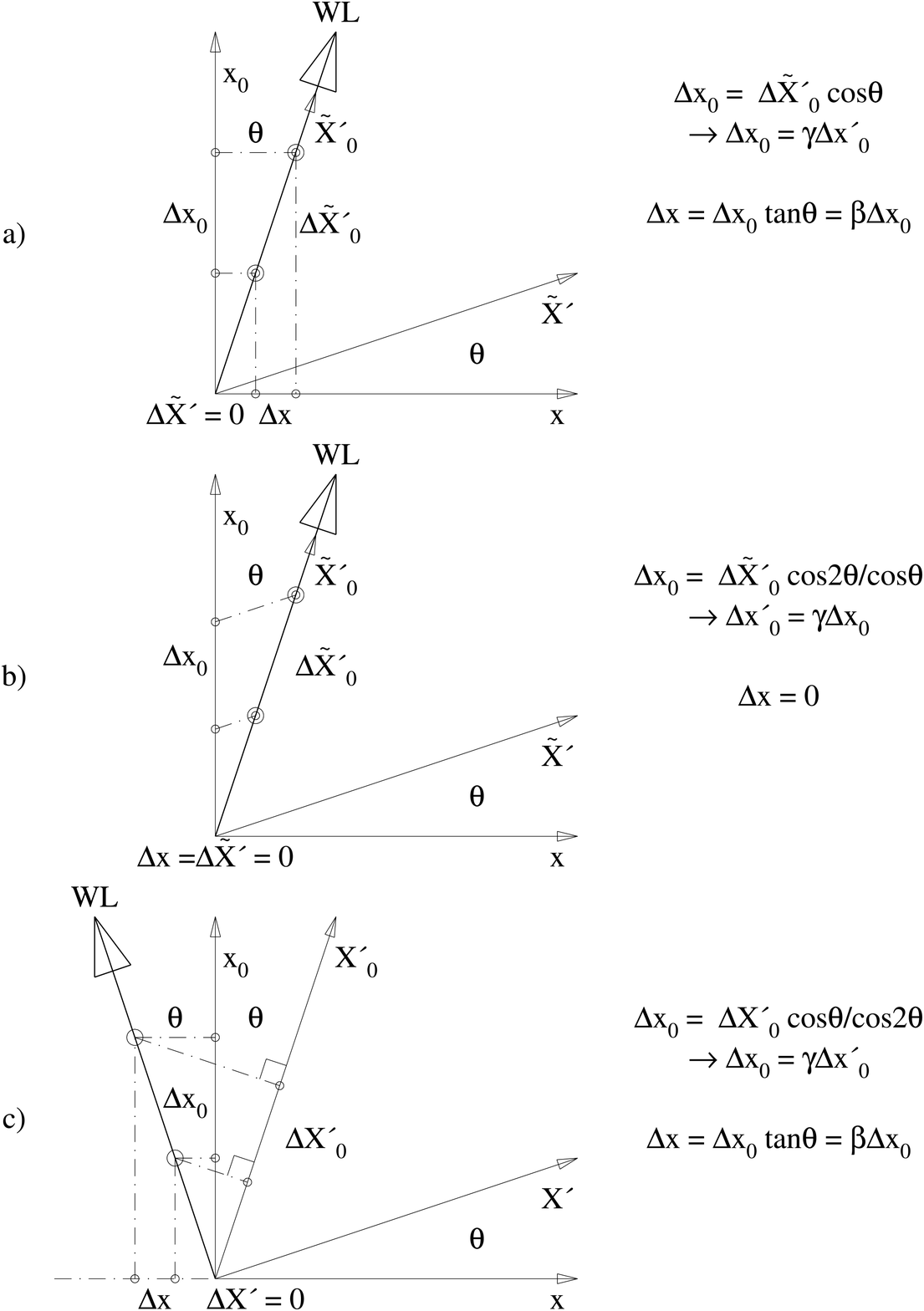}}
\caption{ {\em  Time transformation using different projection procedures: a)
 perpendicular to the $x_0$ axis as in Ref.~\cite{TW}, b) oblique, parallel
 to the $\tilde{X}'$ axis, c) Correct prediction of the LT. Fortuitously,
 a) and c) give the same prediction.}}  
\label{fig-fig10}
\end{center}
\end{figure}

     \par The discussion of `relativity of simultaneity' based on Fig.67 of Ref.~\cite{TW} is
    illustrated in Fig.11. The events 1 and 2 have the same $x'_0$ coordinates, but
    $x2_0 > x1_0$. The events 3 and 4 have the same $x_0$ coordinates but $x'3_0 > x4'_0$.
    A similar argument applied to Fig.9b where the correct projective geometry of the LT is applied
     would seem to indicate that simultaneous events in the frame S at the ends $\sharp$ 1 and
  $\sharp$ 2 of the object considered are not simultaneous in S', since the projections onto the
    $X1'_0$ axis give $X1'_0({\rm end} \sharp 2) > X1'_0({\rm end} \sharp 1)$. Such a `relativity
    of simultaneity' effect would, however, be in direct contradiction to translational invariance
     of the time dilatation relation: $\Delta t = \gamma \Delta t'$, pointed
    out in Section 2 above. To understand this apparent contradiction it is
      necessary to examine more closely
    the operational meaning of `relativity of simultanenty' in an actual experiment. Clearly, at
     least two spatially separated clocks must be introduced into any such discussion, whereas, in
     the examples just discussed, based on Fig.11 or Fig.9, only a single clock --- the one at
     the end $\sharp$ 1 of the object--- is considered. Introducing clocks at both ends of
     the object, synchronised in
      S,(i.e. clocks recording the same time, $t = x_0/c$, at any instant) and denoting
      the corresponding times in the frame S' by $t'_1 =  x1'_0/c$ and $t'_2 =  x2'_0/c$,
      the oblique projection of
      Fig.11  gives the space-time configuration shown in Fig.12a, where
      different scaled time and space coordinates, as in Eqn(5.4) and (5.13), have been introduced for
      end $\sharp$ 1 and end  $\sharp$ 2 of the object. It is clear from the geometry of this
       figure that, due to TD, for any pair of
       events on the world lines of end $\sharp$ 1 and end  $\sharp$ 2 in S for which 
       $\Delta x_0 = 0$, then $\tilde{X}1'_0 = \tilde{X}2'_0$ and hence 
         $x1'_0 = x2'_0$ ---events which are simultaneous in S are also simultaneous in S'---
        there is no `relativity of simultaneity' in this case.
         The incorrect procedure (according to the LT) of obliquely projecting an event on the world line of 
       end $\sharp$ 2 onto the $\tilde{X}1'_0$ axis,  as in Fig.12a, gives the interval
        $\Delta\tilde{X}'_0$, but is not indicative of any  `relativity of simultaneity' of clocks, 
         at the ends of the object, synchronised in the frame S, when observed from the frame S'
          In fact, in Fig.12a, oblique projections of events 1 and 2 onto the
        $\tilde{X}1'_0$ or
        $\tilde{X}2'_0$ axes give the four time intervals  O$_1$M, O$_1$1, O$_2$2 and O$_2$L,
        but only ${\rm O}_11 = \tilde{X}1'_0 = {\rm O}_22 = \tilde{X}2'_0$ represent the
       times of the clocks in S' The spurious `relativity of simultaneity' effect
       of Fig.11 arises
       from falsely identifying either O$_1$M or O$_2$L with corresponding clock times in
       S'.
        \par Similar conclusions are drawn from Fig.12b, analogous to Fig.5, where the correct
        projection procedure relating coordinates in the frames S and S' is used.
        Orthogonal projections of the points 1 and 2 on the world lines of end $\sharp$ 1
        and end $\sharp$ 2 of the object onto either the $X1'_0$ or $X2'_0$ axes define
       the four time intervals O$_1$P, O$_1$Q, O$_2$N and O$_2$R which are
       related by the condition
        \begin{equation}
          \Delta X'_0 = {\rm O}_1 {\rm Q} - {\rm O}_1 {\rm P}
                   = {\rm O}_2 {\rm R} - {\rm O}_2 {\rm N}
       \end{equation}
       However, only two of these intervals, ${\rm O}_1 {\rm P} = X1'_0$ and
       ${\rm O}_2 {\rm R} = X2'_0$ correspond to the times registered by the clocks in S'
       at time $x_0$ in the frame S. The spurious `relativity of simultaneity' effect arises
        from incorrectly identifying either of the time intervals  O$_1$Q or  O$_2$N with the 
       observed time in S' of an event at time $x_0$ in S.

     \begin{figure}[htbp]
\begin{center}\hspace*{-0.5cm}\mbox{
\epsfysize8.0cm\epsffile{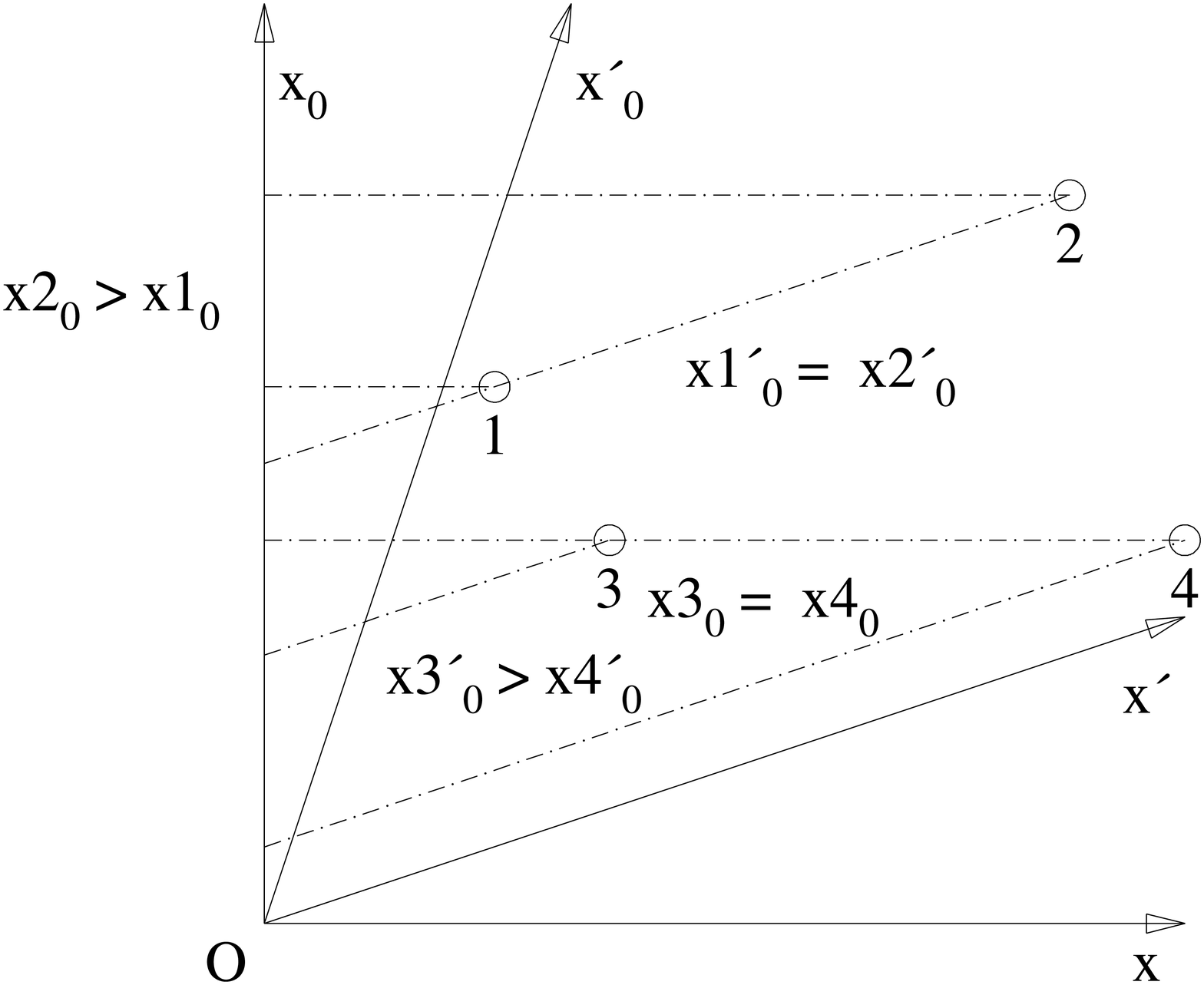}}
\caption{ {\em  Figure showing apparent `relativity of simultaneity' from Ref.~\cite{TW}. See text
   for discussion.}}  
\label{fig-fig11}
\end{center}
\end{figure}

       \par The non-existent `relativity of simultaneity' effect then results, unlike
        `length contraction', not from geometrical errors in the Minkowski plot
       but from a fundamental misunderstanding of the physical meaning of time
       intervals corresponding to certain geometrical projections, regardless
       of whether the latter are correct (as predicted by the LT) as in Fig.12b, or not,
       as in Fig.12a. It is essential that
       two spatially separated and synchronised clocks, each with its own time reading, be introduced
       into the problem in order to perform any meaningful analysis of it.
        This is not the case for Fig.11, or the equivalent Fig.67
       of Ref.~\cite{TW}.

     \begin{figure}[htbp]
\begin{center}\hspace*{-0.5cm}\mbox{
\epsfysize16.0cm\epsffile{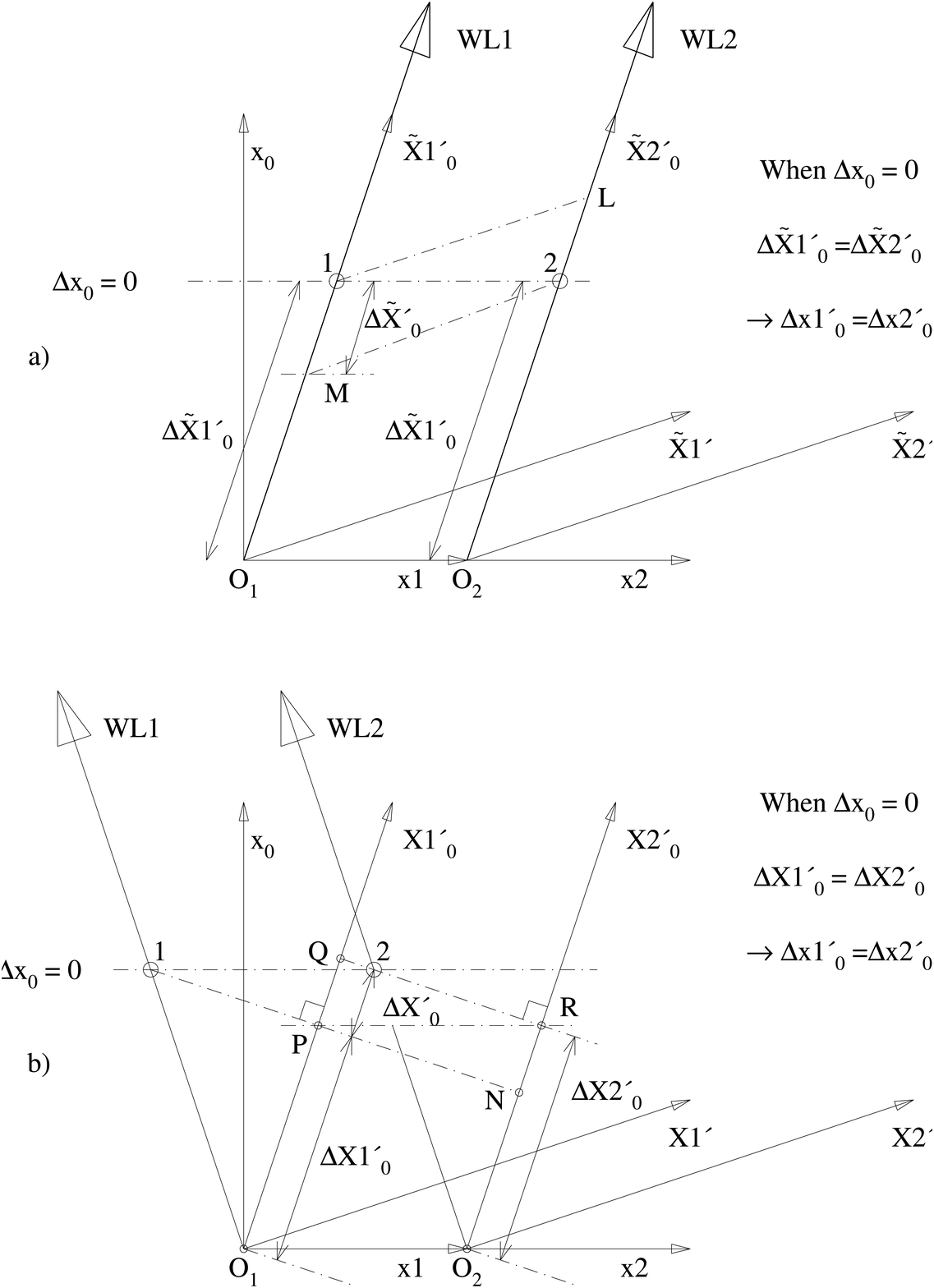}}
\caption{ {\em Discussion of simultaneity introducing the world lines of separate
 clocks at the ends of the object discussed in Ref.~\cite{TW}: a) oblique projections similar
  to that performed on points 3 and 4 in Fig.11, b) correct prediction of the LT. In neither
   case is there any `relativity of simultaneity' effect. See text for discussion.}}  
\label{fig-fig12}
\end{center}
\end{figure}

        \par The discussion of `length contraction' in Mermin's book `Space and Time in Special
             Relativity'~\cite{Mermin} is similar to that of Taylor and Wheeler~\cite{TW}.
           Oblique projections are used and the $t'$ axis is incorrectly drawn, following
          Minkowski, along the world line in S of the origin of the frame S'. The geometry
         of Fig.17.24 of Ref.~\cite{Mermin}, illustrating  Mermin's interpretation
        of `length contraction' is the same as that of Fig.9a of the present paper 
         and the same incorrect result (5.12) is obtained.

     \begin{figure}[htbp]
\begin{center}\hspace*{-0.5cm}\mbox{
\epsfysize16.0cm\epsffile{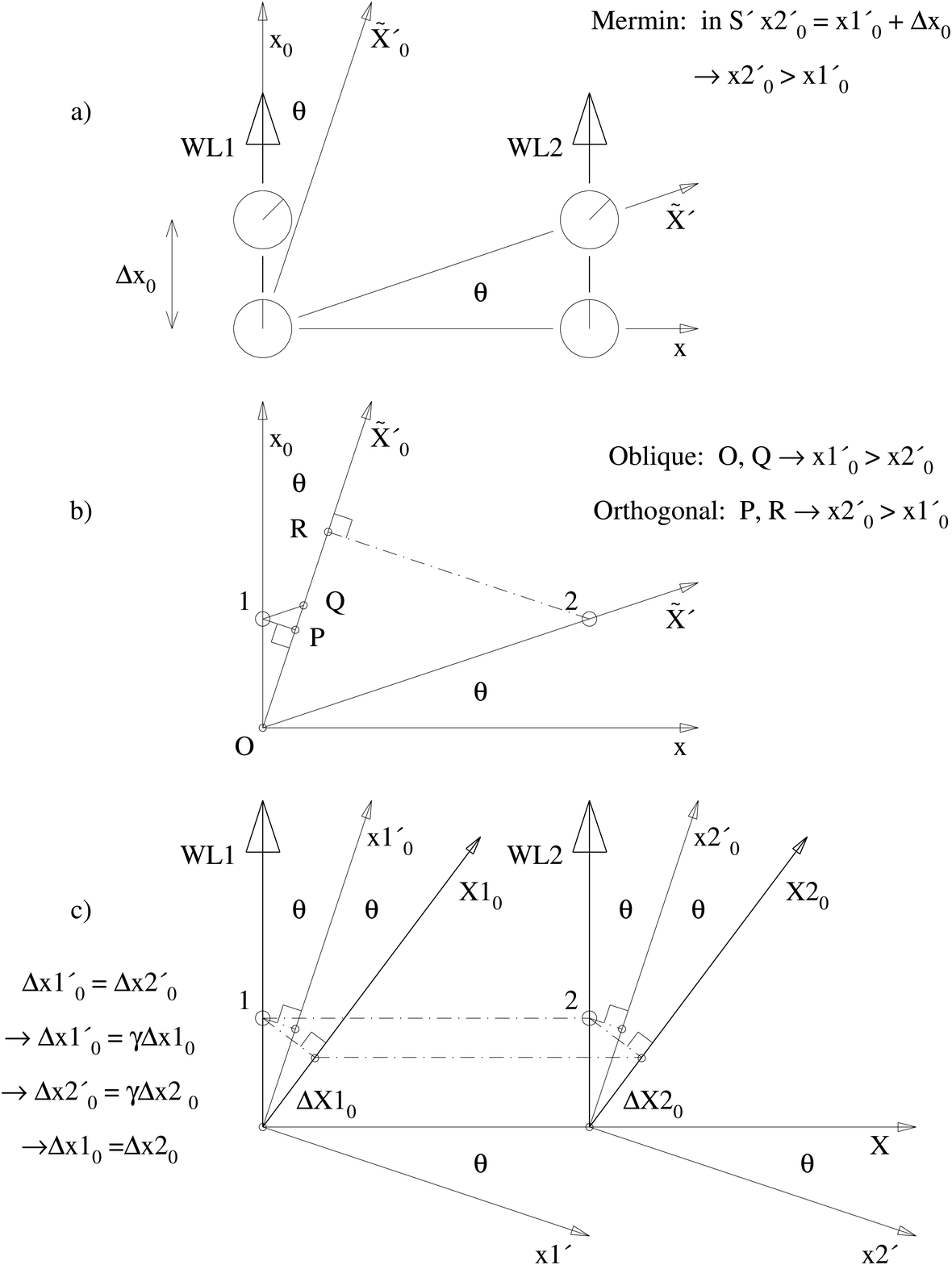}}
\caption{ {\em a) Figure claiming to demonstrate `relativity of simultaneity' from 
 Ref.~\cite{Mermin}, b) Naive `relativity of simultaneity' effects given by oblique or
 perpendicular projections as in Fig.11. c) correct calculation of the time transformation
 corresponding to clocks at rest in S, as shown in a). See text for discusion.}}  
\label{fig-fig13}
\end{center}
\end{figure}

         \par  Mermin's discussion of `relativity of simultaneity' is, however, somewhat 
          different to that of other authors. Instead of considering, as in Ref.~\cite{TW}
          the abstract geometrical projections of Fig.11, the world lines of two clocks at rest in 
         the frame S are considered, which at least, unlike Fig.11, satisifies the minimum
         condition for a meaningful discussion of the problem. In Fig.13a is shown a fair
         copy of Fig.17.22 of Ref.~\cite{Mermin}, the basis of the discussion there
        of `relativity of simultaneity'.
         The S frame coordinates $\tilde{X}'$ and  $\tilde{X}'_0$ related to $x'$ and $x'_0$
         by the relations (5.13) and (5.4), repectively,  also given by Mermin, are shown. The clocks
        are synchronised at $x_0 = 0$ and it is shown that at the later time when the
        world line of clock 2 crosses the  $\tilde{X}'$ axis an observer in S' will see that
        it is in advance of the clock 1 by the time interval, $\Delta x_0$ that
         has elapsed in the frame S. This is supposed to be evident from the figure
         (perhaps from an oblique projection parallel to the $\tilde{X}'$ axis?)
        since no supporting explanation or calculation is given. The claimed
          `relativity of simultaneity' effect is such that $x2'_0 > x1'_0$. Comparing
          in Fig.13b different projections of simultaneous events in S on the world
           lines of the clocks, shows that an oblique projection, as used in Mermin's
            discussion of `length contraction', gives instead $x1'_0 > x2'_0$.
            Projection orthogonal to the $x_0$ axis onto the $\tilde{X}'_0$ axis
              gives simultaneous events in S'. Indeed, Fig.13b indicates that in order to 
           obtain $x2'_0 > x1'_0$ a projection orthogonal to the $\tilde{X}'_0$ axis
            is required. Such a projection is nowhere else mentioned in Ref.~\cite{Mermin}.
          \par The correct simultaneity analysis, according to the LT, of events on the 
           world lines of clocks at rest in S is shown in Fig.13c, which is analogous to
           Figs.5 and 12b. Since the clocks are at rest in the frame S, their world lines
           in the frame S' are: $x1' = -\beta x1'_0$ and $x2' = -\beta x2'_0$.
           The Minkowski plot given by the LT in the frame S' is then the same as Fig.6, but with exchange of primed
           and unprimed coordinates. As can be seen in Fig.13c, The $x'$ and $x'_0$
           axes are orthogonal as a conseqence of the relations, the inverse of
           (2.7) and (2.8):
  \begin{eqnarray}
   X & = & x' \cos \theta+ x'_0 \sin \theta  \\
  X_0 & = & x'_0 \cos \theta+ x' \sin \theta
  \end{eqnarray}
            while the $X$ and $X_0$ axes are obliquely oriented. As evident, in any case, from
      translational invariance, the projections in Fig.13c show that there is no
       `relativity of simultaneity' effect in the problem. Notice that, in this case, the TD effect is inverted
       with respect to that shown in Fig.2 above; i.e. the clocks in the frame S' appear, to an observer in this
      frame, to be running faster than those in S. The space-time experiment shown in Fig.13c is thus the
     reciprocal of that in Fig.2.   

 \SECTION{\bf{Summary}}
  The manifest translational invariance of the TD relations in (2.13) demonstrate
  the spurious nature of correlated `relativity of simultaneity' and `length  contraction'
   effects that have hitherto, following Einstein~\cite{Ein1}, been derived from the LT,
    as explained in previous papers by the present author~\cite{JHFSSC,JHFCRCS,JHFLTASC}
    and Section 2 of the present paper.
  \par Alternative derivations of `length  contraction' from the geometry of the Minkowski plot
   are shown to be flawed by a sign error in an angle, when drawing the $x'$ and $t'$ axes on
   the original
    Minkowski plot~\cite{Mink}, which has been propagated, uncorrected, in essentially 
    all text-book treatments of the subject, as well as the use projection operations
     at variance with those required by the LT. When the correct
    projection and scaling operations derived in Section 3, directly from the LT, are applied, it is
     clear, by simple inspection, (see Fig. 3 and  Fig. 9b) that there is no `length  contraction' effect.
    \par The `relativity of simultaneity' effect derived from the Minkowski plot, unlike that
    obtained in Eqns(2.14)-(2.17), directly from misuse of the LT, is not a direct consequence of the
     `length  contraction' effect obtained from the same plot. It is, instead, produced by a false 
    identification of certain projections on the plot with the times recorded by two
    spatially separated and synchronised clocks. It is shown in Fig.12a, that, even with an incorrect
    oblique projection procedure, similar to that used to obtain `length contraction', there is no corresponding
     `relativity of simultaneity' effect. 
   \par The present author has been able to find only one text book in which the $x'$ and $t'$ axes of
    the Minkowski plot are correctly drawn~\cite{Anderson}. However, the plot is not used in the book
    in the discussion of the `length contraction' effect.

\pagebreak

\end{document}